\documentclass[prb,aps,floatfix,twocolumn,showpacs]{revtex4}
\usepackage{eurosym}
\usepackage{graphicx,graphics,color,epsfig}
\usepackage{amsmath}
\usepackage{color}
\usepackage{amsfonts}
\usepackage{amssymb}
\usepackage{bm}

\begin{document}

\title{Valley Splitting and Valley Dependent Inter-Landau-Level Optical
Transitions in Monolayer MoS$_{2}$ Quantum Hall Systems}
\author{Rui-Lin Chu$^1$}
\author{Xiao Li$^2$}
\author{Sanfeng Wu$^{3}$}
\author{Qian Niu$^2$}
\author{Wang Yao$^{4}$}
\author{Xiaodong Xu$^{3,5}$}
\author{Chuanwei Zhang$^{1}$}
\email{chuanwei.zhang@utdallas.edu}
\affiliation{$^{1}$ Department of Physics, the University of Texas at Dallas,
Richardson,TX 75080 USA}
\affiliation{$^{2}$Department of Physics, The University of Texas at Austin, Austin,
Texas 78712, USA}
\affiliation{$^{3}$Department of Physics, University of Washington, Seattle, Washington,
USA}
\affiliation{$^{4}$Department of Physics and Center of Theoretical and Computational
Physics, University of Hong Kong, Hong Kong, China}
\affiliation{$^{5}$Department of Material Science and Engineering, University of
Washington, Seattle, Washington, USA}

\begin{abstract}
The valley dependent optical selection rules in recently discovered
monolayer group-VI transition metal dichalcogenides (TMDs) make possible
optical control of valley polarization, a crucial step towards valleytronic
applications. However, in the presence of Landaul level (LL) quantization
such selection rules are taken over by selection rules between the LLs,
which are not necessarily valley contrasting. Using MoS$_{2}$ as an example
we show that the spatial inversion-symmetry breaking results in unusual
valley dependent inter-LL selection rules, which is controlled by the sign
of the magnetic field and directly locks polarization to valley. We find a
systematic valley splitting for all LLs in the quantum Hall regime, whose
magnitude is linearly proportional to the magnetic field and in comparable
with the LL spacing. Consequently, unique plateau structures are found in
the optical Hall conductivity, which can be measured by the magneto-optical
Faraday rotations.
\end{abstract}

\pacs{73.43.-f, 71.70.Di, 78.55.Ap, 71.35.Ji}
\maketitle

Optical properties of two-dimensional (2D) charge carrier systems such as 2D
electron gases (2DEG), graphene, and topological insulators are important
for studying their underlying charge carrier properties and future
applications in optoelectronics \cite%
{Klem,IVK,Kerr,Ploog,Ploog2,deHeer,Carbotte,Jiangzg,Aguilar,Tse,Tokura,Abergel}%
. Generally, the charge carrier dynamics can be strikingly different with
and without a magnetic field, as evidenced by the celebrated example of
quantum Hall effect (QHE). In the quantum Hall regime, an external large
magnetic field produces a series of Landau levels (LLs) with discrete
energies and the optical transitions occur only between appropriate LLs
following certain selection rules. In the past few decades, such selection
rules and relevant optical phenomena have been intensively studied in
various magneto-optical measurements \cite%
{Klem,IVK,Kerr,Ploog,Ploog2,deHeer,Jiangzg,Aguilar,Tokura,Abergel,Iris,Shimano,Hofmann}%
.

Monolayers of MoS$_{2}$ and other group-VI transition-metal dichalcogenides
(TMDs) represent a new family of 2D materials beyond graphene. Because of
their coupled spin and valley physics and large band gap, monolayer TMDs
have become exciting platforms for exploring novel valleytronic and
optoelectronic applications \cite%
{Mak-10prl,zhuzhiyong,AKis,xiaodi-mo,nat-com,Mak2, xiaodongcui}. Recently,
the optical properties of TMDs in zero magnetic field have been widely
studied in many experiments \cite{xiaodongcui,Mak2,xu1,xu2}. However, the
counterpart in a finite magnetic field has not been well explored.

In this work, we study the optical properties of monolayer MoS$_{2}$ quantum
Hall systems with a large magnetic field. With a zero magnetic field, it is
known that monolayer MoS$_{2}$ and graphene share similar valley contrasting
physics, i.e., the circular polarizations ($\sigma _{+}$, $\sigma _{-}$) are
locked with two inequivalent valleys K and K$^{\prime }$ due to the opposite
orbital helicity of two valleys \cite{Geim,yaowang,beenakker,white}.
However, with a large magnetic field in graphene, the polarization is no
longer associated with the valley degree of freedom because the transitions
are between LLs, whose selection rules allow transitions in both
polarizations ($\sigma _{+}$ and $\sigma _{-}$) at both valleys (K and K$%
^{\prime }$), as observed in many magneto-optical measurements \cite%
{deHeer,Jiangzg,Iris,Shimano,Hofmann}. Furthermore, such inter-LL selection
rules in graphene are the same as 2DEG with large band gaps (e.g., GaAs
quantum wells), where valleys do not even exist \cite%
{Klem,IVK,Kerr,Ploog,Ploog2}. These known results naturally indicate that
the optical responses of monolayer MoS$_{2}$ in the quantum Hall region
should also be valley independent.

Surprisingly, we find this is not the case for monolayer MoS$_{2}$, where
the polarization selection rules for the inter-LL transitions are still
valley dependent. More interestingly, the selection rules are controlled by
the sign of the magnetic field, i.e., the valley index in the selection
rules can be flipped by reversing the sign of the magnetic field, which is
fundamentally different from the zero magnetic field case, where the
polarization-valley locking is fixed. We also show that optical transitions
in this system are made more unusual by a systematic valley splitting for
all LLs, whose magnitude is linear against the magnetic field and is
comparable with the LL spacing. The valley-polarization selection rules and
valley splitting lead to a series of spin-valley polarized transitions
between LLs as well as unique plateau structures in the optical Hall
conductivity, which can be addressed in the circular dichroism,
magnetoluminescence and Faraday rotations, showing distinguishable features
from graphene and 2DEG. Our predictions also apply to other group-VI TMDs.

\textit{LLs and valley splitting.} The monolayers of MoS$_{2}$ consist of a
Mo layer sandwiched between two S layers in a trigonal prismatic
arrangement. Although similar to graphene in many aspects, some of its
properties are more favorable than graphene. It features a direct band gap $%
\Delta $ in the visible wavelength regime, which occurs at the two
inequivalent valleys K and K$^{\prime }$ at the corners of the hexagonal
Brillouin zone. The inversion symmetry is naturally broken in the
monolayers, which induces both strong spin-orbit coupling and spin-valley
coupling \cite{xiaodi-mo,nat-com}.

\begin{figure}[t]
\centering
\includegraphics[width=0.48\textwidth]{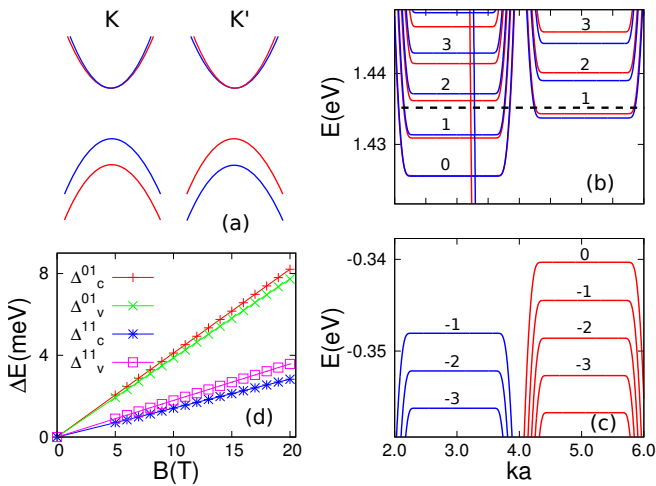}\newline
\vspace{-6pt}
\caption{ (color online) (a) schematic of the spin-valley coupled band
structure of TMDs. Red(blue) represents spin up(down), respectively. (b) and
(c): Conduction and valence band LLs for MoS$_{2}$ under $B_{\perp }$ = 20
T. K(K') valley is on the left(right). The crossing-LL states in the
conduction band are from the dangling bonds on the zigzag edges\protect\cite%
{Liuguibin}, which do not affect the LLs. Dash line is a guide to eye for
valley spitting and also marks filling level $\protect\nu(K)=4$, $\protect\nu%
(K^{\prime })=2$. (d) $\Delta _{c}^{01}$($\Delta _{v}^{01})$ is the absolute
energy difference between the LL 0(-1) in K valley and 1(0) in K' valley in
conduction(valence) band. $\hbar\protect\omega_0^c$($\hbar\protect\omega_0^v$%
) is the LL spacing in conduction(valence) band. (e) same as (d) calculated
from orbital magnetic moment and effective mass approximation. $a$ is the
lattice constant.}
\end{figure}

The zero-field band structure for monolayer MoS$_{2}$ is schematically shown
in Fig. 1(a), which is usually described by the effective Dirac model \cite%
{xiaodi-mo,nat-com}. In each valley there are two sets of gapped Dirac
spectra with red (blue) representing spin up (down) respectively. Because of
the large effective mass at the band edges the LLs only scale as $n\hbar
eB_{\perp }/m^{\ast }$ at the low energy part, which resemble conventional
2D semiconductors more rather than Dirac fermions \cite%
{Lixiao,haizhou,Asgari,Andor-prb,Zhouli}. Here $B_{\perp }$ is the
perpendicular magnetic field and $n$ is the LL index. To obtain the LLs, we
adopt a three-band atomic tight-binding model from Ref. \cite{Liuguibin} and
apply $B_{\perp }$ via the Peierls substitution $t_{ij}=t_{1,2}e^{-ie/\hbar
\int \mathbf{A}\cdot d\mathbf{r}}$, where $\mathbf{A}=(-By,0,0)$ is the
vector potential.

In Fig. 1(b-c) we present the low energy LLs, where a zigzag ribbon
structure is used with a width $L_{y}=170$ nm, which is sufficiently larger
than the magnetic length scale $l_{B}$. The set of LLs from the lower
split-off valence bands are not shown since they are similar to Fig. 1(c).
The Zeeman splitting is first neglected here and will be discussed later. We
label the LLs and assign the $n=0$ LLs according to the analytic solutions
from the effective two-band Dirac model \cite{Lixiao}. When $B_{\perp }>0$
they appear only in conduction band of K valley and valence band of K$%
^{\prime }$ valley. Therefore the valley degeneracy for them is already
lifted. Here our focus is on the more general $n\neq 0$ LLs.

We notice a systematic valley splitting exists for all $n\neq 0$ LLs with
the magnitude comparable to the LL spacing, which is not revealed by the
effective model \cite{Lixiao}. A linear relation with $B_{\perp }$ is found.
Here we let $B_{\perp }\geqslant 5$T to ensure $L_{y}>>l_{B}$. The linearity
should extend to the low field situation in this single-particle
calculation. The linear valley splitting and its discrepancy with the
effective model can be intuitively understood from the orbital magnetic
moment \cite{Chang,prx-dot}, 
which is of opposite sign at the two valleys ($\pm m$). Taking the
conduction band as an example, in the presence of $B_{\perp }$ the valley
energy difference is $\Delta _{c}^{01}=2m\cdot B_{\perp }$. In the effective
model, this matches $\hbar \omega _{0}^{c}$, the LL spacing between 0 and 1,
resulting in valley degeneracy in $n\neq 0$ LLs \cite{Cai}. Such matching is
violated in the tight-binding model where $\Delta _{c}^{01}>\hbar \omega
_{0}^{c}$ giving rise to the valley splitting, as is shown in Fig. 1d and
1e. In the case of graphene, the valley degeneracy is known to be lifted in
high magnetic fields via electron-electron or electron-phonon interactions 
\cite{Kim-prl,Ong-prl,Andrei-nat,Kim-natphys}. Similar linear relations
between the valley splitting and $B_{\perp }$ have also been experimentally
observed in silicon and AlAs 2D electron systems \cite{Si-natphy, AlAs-prl}.
Their physical origin, however, remains controversial.

The valley splitting here has a few direct consequences: (i) The $n=0,1$ LLs
in conduction band are always valley polarized and $n=0,-1$ in valence band
are spin-valley polarized. (ii) The total filling factor follows a sequence $%
\nu =2,3,4\ldots $ in the electron doped regime and $\nu =-1,-2,-3\ldots $
in the hole doped regime. The lifting of valley degeneracy in $n=0$ LLs can
be attributed to the broken spatial-inversion symmetry in the monolayer.
Such symmetry is known to guarantee the valley degeneracy rigorously on
graphene, regardless of the time-reversal symmetry \cite{Mccann}. However,
it does not explain the splitting in $n\neq 0$ LLs \cite{Fuchs}. Instead,
using graphene lattice as a toy model, we find the splitting in $n\neq 0$
LLs can be induced by the next-nearest-neighbour (NNN) electron hopping,
which breaks the electron-hole symmetry. Therefore the valley-splitting in $%
n\neq 0$ LLs stems from spontaneous breaking of spatial-inversion,
electron-hole and time-reversal symmetry. In fact, the low energy physics in
MoS$_{2}$ is dominated by electron hopping between Mo atoms, which is indeed
the NNN hopping on the honeycomb lattice \cite{xiaodi-mo,nat-com}.

\begin{figure}[t]
\centering
\includegraphics[width=0.47\textwidth]{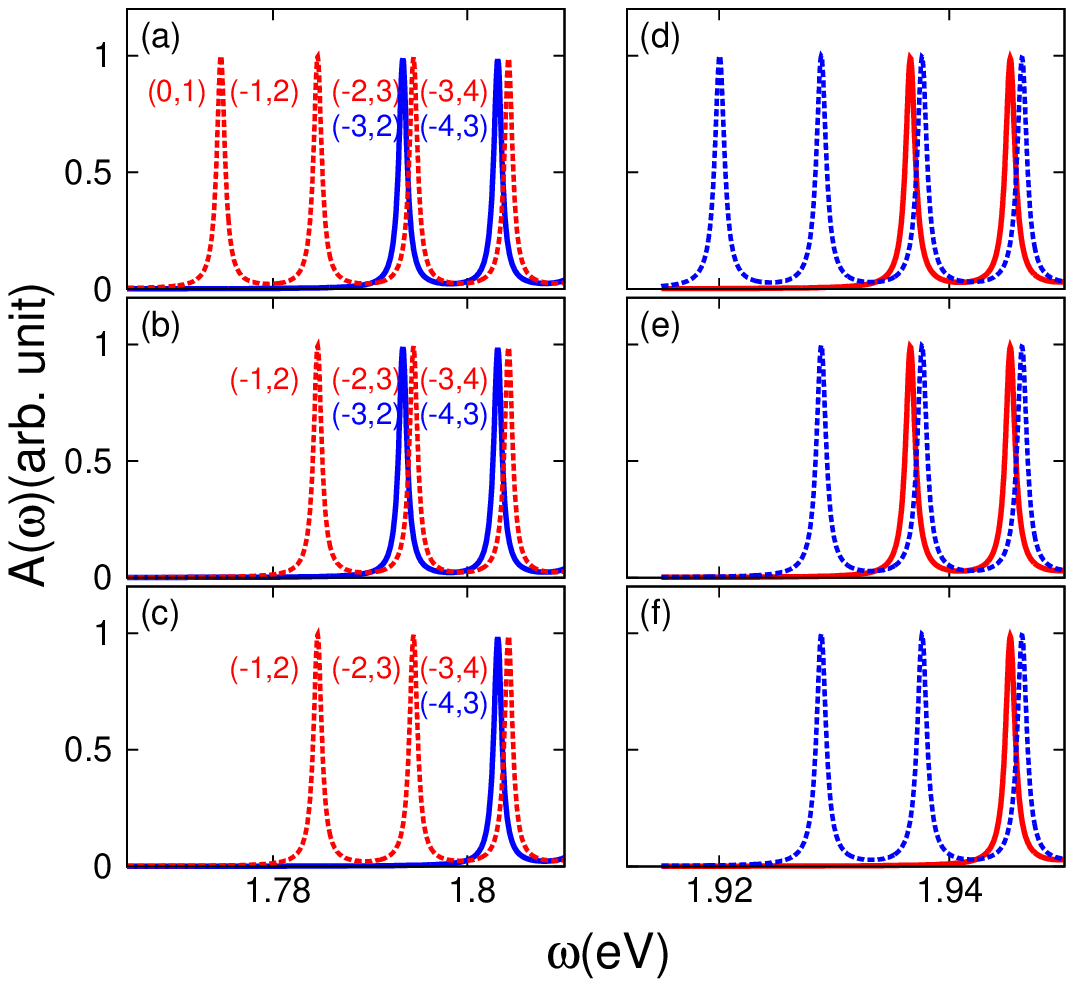}\newline
\includegraphics[width=0.4\textwidth]{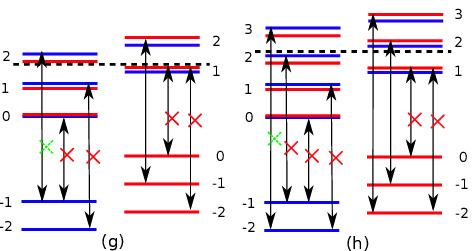}\newline
\vspace{-4pt}
\caption{ (color online) Optical absorption spectrum in the quantum Hall
regime. The solid(dash) lines represent K(K') valleys, respectively.
Red(blue) represents spin up(down), respectively. Spin-valley polarized
transitions between $(n,n^{\prime })$ are labeled. (a)\&(d): $\protect\nu %
(K)=4$, $\protect\nu (K^{\prime })=0$. (b)\&(e): $\protect\nu (K)=4$, $%
\protect\nu (K^{\prime })=2$. (c)\&(f): $\protect\nu (K)=6$, $\protect\nu %
(K^{\prime })=2$. $\Gamma =0.1\hbar \protect\omega _{0}$. (g-h) Schematic of
the inter-LL transitions corresponding to (b,e) and (c,f). Red(solid)
crosses indicate Pauli blocking. Green(dash) crosses indicate vanishing
probability. }
\end{figure}

\textit{Valley-dependent inter-LL selection rules:} We now turn to the
optical properties of these valley-degeneracy-lifted LLs. Because $\hbar
\omega _{0}<<\Delta $, the intraband and interband optical transitions in
this system belong to two completely different regimes: intraband in the
microwave to terahertz and interband in the visible frequency range. We will
set our focus on the latter because for MoS$_{2}$ the valley contrasting
interband optical transitions have been the most intriguing property in
experiments for valleytronics \cite%
{xiaodi-mo,nat-com,xiaodongcui,Mak2,xu1,xu2}.

When considering the transitions between levels $n^{\prime }$ and $n$, the
well-known selection rule for 2DEG and graphene requires $|n|=|n^{\prime
}|\pm 1$ \cite{Aoki-prl,Carbotte, Jiangzg, Aoki-prb}. For MoS$_2$ such
selection rule can also be obtained from the effective model\cite{Rose}. At
first glance, since the LL spacing is comparable in the conduction and
valence bands, four transitions would occur at very close but non-degenerate
photon energies: $-n\leftrightarrow n+1$ and $-(n+1)\leftrightarrow n$ ($%
n\geqslant 1$) for both valleys. In Fig. 2 we calculate the optical
absorption spectrum 
\begin{equation}
A(\omega )\propto \sum_{\epsilon _{n}<\mu ,\epsilon _{n^{\prime }}>\mu }%
\frac{|J_{x}^{nn^{\prime }}|^{2}}{i(\epsilon _{n^{\prime }}-\epsilon
_{n}-\omega -i\Gamma )},
\end{equation}%
at three different filling levels corresponding to the LLs presented in
Fig.1 through exact diagonalization, where $J_{x}$ is the current matrix and 
$\Gamma $ is the broadening parameter. We immediately notice several
distinctive features. (i) The expected four-fold peaks only appear two-fold.
Unlike in graphene, here the transitions $-n\leftrightarrow n+1$ in K valley
and $-(n+1)\leftrightarrow n$ in K$^{\prime }$ valley are completely
suppressed, which suggest highly valley dependent selection rules.
Transitions from the spin-split lower valence bands at higher photon
energies also follow the same rule except with opposite spins, as seen in
Figs. 2(d-f). Such selection rules originate from the severely broken
spatial-inversion symmetry in the monolayer. (ii) As the filling level goes
up, the number of spin-valley polarized peaks has an alternating $2-1-2-1$
pattern, which can be attributed to the valley-imbalanced Pauli blocking
caused by the $n=0$ LLs and the valley splitting as illustrated in Figs.
2(g-h).

\begin{figure}[t]
\centering
\includegraphics[width=0.33\textwidth]{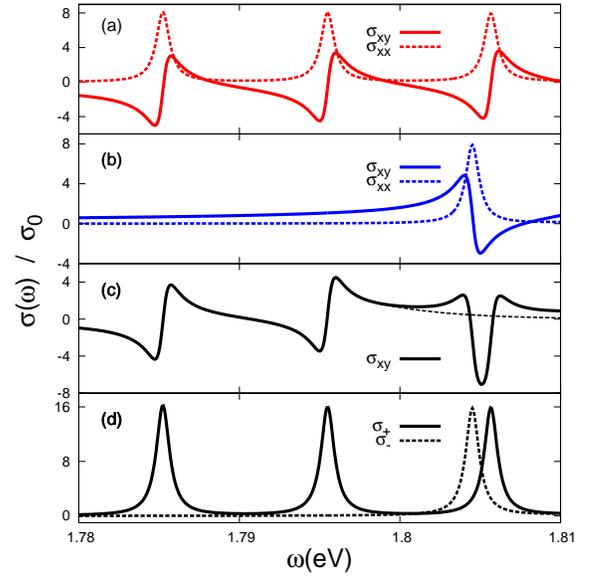}\newline
\vspace{1pt}
\caption{ (color online) Optical Hall conductivity in unit of $\protect%
\sigma _{0}=e^{2}/h$ with filling factors $\protect\nu (K)=6$, $\protect\nu %
(K^{\prime })=2$, corresponding to Fig. 2(c). (a): optical conductivity $%
\protect\sigma _{xy}$ and $\protect\sigma _{xx}$ for spin up, K$^{\prime }$
valley. (b): spin down, K valley. (c) total $\protect\sigma _{xy}$, the dash
line schematically shows the cancelling out situation when $\Delta
_{c}^{01}=\Delta _{v}^{01}$. (d): total $\protect\sigma _{+/-}$. Only the
real part is plotted.}
\end{figure}

To further understand the role played by the valley degree of freedom in the
optical Hall effect, we calculate the optical Hall conductivity using the
Kubo formula \cite{Aoki-prl,Aoki-prb,Shimano}, 
\begin{eqnarray}
\sigma _{ij}(\omega ) &=&\frac{i}{L_{x}L_{y}}\sum_{\epsilon _{n}<\mu
,\epsilon _{n^{\prime }}>\mu }\frac{1}{\epsilon _{n^{\prime }}-\epsilon _{n}}%
(\frac{J_{i}^{nn^{\prime }}J_{j}^{n^{\prime }n}}{\epsilon _{n^{\prime
}}-\epsilon _{n}-\omega -i\Gamma }  \notag \\
&-&\frac{J_{j}^{nn^{\prime }}J_{i}^{n^{\prime }n}}{\epsilon _{n^{\prime
}}-\epsilon _{n}+\omega +i\Gamma }),
\end{eqnarray}%
where $i,j=x,y$. Following Ref. \cite{Aoki-prl}, here we retain 40 LLs and
impose periodic boundary conditions along the $x$ and $y$ directions. $%
B_{\perp }$ is kept at $20.8$ T, as close as possible to that used Figs. 1
and 2, since in such calculations $B_{\perp }$ can only take discrete
levels. The two valleys cannot be distinguished in the momentum space.
However, since the valley index is associated with spin, we can distinguish
valleys by spins. The result is shown in Fig. 3, where $\sigma _{\pm
}(\omega )=\sigma _{xx}(\omega )\pm i\sigma _{xy}(\omega )$ is the optical
conductivity for the right and left circular polarized light. Spin-valley
polarized resonance structures for $\sigma _{xy}$ is found, similar to Fig.
2. In this set-up, for the electron(hole) doped regime the spin-valley
polarization is achieved for the spin up (down) and K$^{\prime }$ (K)
valley, respectively. Upon switching the sign of $B_{\perp }$ the valley and
spin polarization also flips. At resonance photon frequencies $\sigma _{xy}$
from the two valleys actually have the opposite signs. This is another
distinctive feature from graphene, in which both valleys contribute equally
to the total $\sigma _{xy}$ \cite{Iris,Shimano}. But due to the difference
in $\Delta _{c}^{01}$ and $\Delta _{v}^{01}$ (Fig. 1d), the resonance
frequencies in the two valleys are slightly miss-matched, leading to
spin-valley mixed resonance peaks in $\sigma _{xy}$ (starting from the third
resonance in Fig. 3c) instead of cancelling out.

\begin{figure}[t]
\centering
\includegraphics[width=0.4\textwidth]{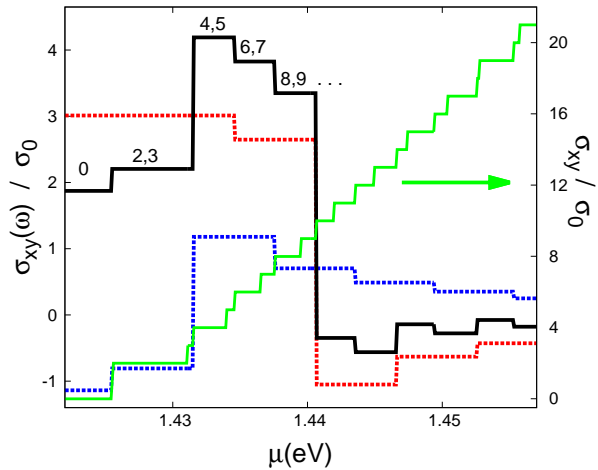}\newline
\vspace{-9pt}
\caption{ (color online) Plateau structure for the optical Hall conductivity 
$\protect\sigma _{xy}$ (black solid line) at $\protect\omega =1.786$ eV and
the static quantum Hall conductivity (green solid line) in the electron
doped regime. Red(blue) dash line represents spin up(down) or K'(K) valley
component of $\protect\sigma _{xy}$, respectively. Numbers on the solid
black line indicate the corresponding quantum Hall conductivity.}
\end{figure}

A more important message from Fig. 3 when compared with Fig. 2 is that the
allowed interband transitions in K and K$^{\prime }$ valleys are solely
attributed to the left and right circular polarized light separately 
\begin{eqnarray}
&&K:~~~-(n+1)~\leftrightarrow ~n,~~~\sigma _{-}  \notag \\
&&K^{\prime }:~~~-n~\leftrightarrow ~n+1,~~~\sigma _{+}
\end{eqnarray}%
where $n\geqslant 0$. Consequently the circular polarization is directly
locked with the valley degree of freedom in optical transitions in the
quantum Hall regime. Upon flipping the direction of $B_\perp$, the valley
index will switch, \textit{i.e.} $K$($\sigma_+$) and $K^{\prime }$($\sigma
_- $).

\textit{Optical Hall plateaus.} The optical conductivity as a function of
the chemical potential $\mu $ is shown in Fig. 4, where $\omega $ is
slightly away from resonance. The static quantum Hall conductivity is also
presented, showing fully valley-degeneracy-lifted and well quantized
plateaus. We notice that in the optical conductivity each spin or valley
component also develops its own and contrasting plateaus, although like the
net $\sigma _{xy}(\omega )$ they are not quantized either. The $n=0$ LLs and
the valley splitting are manifested in the alternating sequence of the step
structures in these two components, which also lead to a unique sequence of
filling factors in the net $\sigma _{xy}(\omega )$ plateaus, as labelled in
Fig. 4. To be specific, $2,3$ spans $n=0$ to $n=1$ in K valley and $4,5$ $%
n=1 $ in K to the $n=1$ in K$^{\prime }$ valley, and so on. Interestingly,
the valley contrasting plateaus persist even when $\mu $ is in the band gap,
as is seen for the $0$ plateau that extends all the way to the valence band
top.

\textit{Circular dichroism, magnetoluminescence and Faraday rotation.}
Valley resolved interband optical transitions shown in Fig. 2 are readily
detectable by the circular dichroism spectroscopy due to the
polarization-valley locking. Given the already excellent photoluminescenence
of monolayer TMDs in zero-field, magnetoluminescence would be an ideal test
of the valley dependent selection rules, in which luminescence between
individual LLs in a selected valley can be driven by resonant circular
polarized excitations in the Faraday geometry \cite{Klem,Ploog,Ploog2}. The
optical Hall conductivity $\sigma _{xy}(\omega )$ can be measured from the
Faraday rotation angle $\theta (\omega )$ \cite{Iris,Shimano,Aoki-prl}. A
spin-valley polarized excitation would be indicated by a single maximum
absolute slope $|d\theta /d\omega |$ at resonant frequencies as shown in
Fig. 3c.

\textit{Interplay of valley and spin splitting.} The Zeeman splitting is
estimated to be much smaller than the valley splitting when we assume an
ordinary g-factor for electrons ($g=2$). The Zeeman spin splitting will
slightly enlarge the valley spitting in both the valence and conduction
bands with the same magnitude. However, The spin in optical transitions is
conserved and the spin-Zeeman field does not change valley-dependent
inter-LL transition frequency and selection rules. Additionally, the valence
band tops at the K(K') valleys are composed of $m=-2$($m=$2) $d$-orbitals
from the Mo atoms\cite{xiaodi-mo,Liuguibin}, which induce an additional
Zeeman-type splitting term in presence of $B_{\perp }$. The valley splitting
in the valence band is further enlarged by this term. The conduction band
bottoms are not affected by this term because they are composed of the $m=0$ 
$d$-orbitals of Mo atom\cite{xiaodi-mo,Liuguibin}. Accordingly, this
additional term induces valley-contrasting frequency shift for the inter-LL transitions at each valley, which will appear as a linear shift against $B_{\perp }$ for each valley in the spectrums of circular dichroism and magnetoluminescence. The
valley-dependent inter-LL selection rules remain intact.

To conclude, we have shown that valley splitting exists for all the LLs in
monolayer MoS$_{2}$ and other TMDs even without considering the interaction
effects. Optical transitions in the quantum Hall regime follow valley
dependent selection rules controlled by the sign of the magnetic field,
which lock the circular polarization and valley degree of freedom together.
Finally we propose circular dichroism spectroscopy, magnetoluminescence and
Faraday rotation measurements as potential tests for the selection rules as
well as the valley splitting. An interesting extension of current study
would be the disorder and localization effect in the optical Hall effect,
since in this system the mixing of valleys inevitably involves spin
flipping, which is distinct from graphene \cite{haizhou,Aoki-prl}.

\emph{Acknowledgements} R.C. and C.Z. are supported by ARO
(W911NF-12-1-0334), AFOSR (FA9550-13-1-0045), and NSF-PHY (1249293). X.L.
and Q.N. are supported by DOE-DMSE (DE-FG03-02ER45958), NBRPC
(2012CB-921300), NSFC (91121004), and the Welch Foundation (F-1255). S. W.
and X. X. are supported by US DoE, BES, Division of Materials Sciences and
Engineering (DE-SC0008145). W. Y is supported by the Croucher
Foundation(Croucher Innovation Award), and the RGC and UGC of Hong
Kong(HKU705513P,HKU9/CRF/13G,AoE/P-04/08).


\begin{thebibliography}{99}
\bibitem{Klem} S. K. Lyo, E. D. Jones, and J. F. Klem, Phys. Rev. Lett. 61,
2265 (1988). 

\bibitem{IVK} I. V. Kukushkin, K. V. Klitzing, and K. Ploog, Phys. Rev. B
37, 8509 (1988).

\bibitem{Kerr} S. R. Andrews, A. S. Plaut, R. T. Harley, and T. M. Kerr,
Phys. Rev. 8 41, 5040 (1990).

\bibitem{Ploog} S. I. Gubarev, T. Ruf, M. Cardona, and K. Ploog, Phys. Rev.
B 48, 1647 (1993). 

\bibitem{Ploog2} I. V. Kukushkin, R. J. Haug, K. von Klitzing, and K. Ploog,
Phys. Rev. Lett. 72, 736 (1994). 

\bibitem{deHeer} M. L. Sadowski, G. Martinez, M. Potemski, C. Berger, and W.
A. deHeer, Phys. Rev. Lett. 97, 266405 (2006).

\bibitem{Carbotte} V. P. Gusynin, S. G. Sharapov, and J. P. Carbotte, Phys.
Rev. Lett. 98, 157402 (2007).

\bibitem{Jiangzg} Z. Jiang \textit{et al}, Phys. Rev. Lett. 98, 197403
(2007).

\bibitem{Aguilar} R. V. Aguilar \textit{et al}, Phys. Rev. Lett. 108, 087403
(2012).

\bibitem{Tse} W. - K. Tse and A. H. MacDonald, Phys. Rev. B 82, 161104(R)
(2010).

\bibitem{Tokura} S. Bordacs \textit{et al}, Phys. Rev. Lett. 111, 166403
(2013).

\bibitem{Abergel} D. S. L. Abergel and Vladimir I. Fal'ko, Phys. Rev. B 75,
155430 (2007). 

\bibitem{Iris} I. Crassee \textit{et al}, Nature Physics 7, 48 (2011). 

\bibitem{Shimano} R. Shimano \textit{et al}, Nature Communications 4, 1841
(2013). 

\bibitem{Hofmann} P. Kuhne \textit{et al}, Phys. Rev. Lett. 111, 077402
(2013).

\bibitem{Mak-10prl} K. F. Mak, C. Lee, J. Hone, J. Shan, \& T. F. Heinz, 
\textit{Phys. Rev. Lett.} \textbf{105}, 136805 (2010).

\bibitem{zhuzhiyong} Z. Y. Zhu, Y. C. Cheng, and U. Schwingenschlogl, Phys.
Rev. B 84, 153402 (2011).

\bibitem{AKis} B. Radisavljevic, A. Radenovic, J. Brivio, V. Giacometti, \&
A. Kis, \textit{Nature Nanotech.} \textbf{6}, 147--150 (2011).

\bibitem{xiaodi-mo} D. Xiao, G. Liu, W. Feng, X. Xu, and W. Yao, Phys. Rev.
Lett. 108,196802 (2012).

\bibitem{nat-com} T. Cao et al. Nature Communications 3, 887 (2012).

\bibitem{Mak2} K. F. Mak, K. He, J. Shan, \& T. F. Heinz, \textit{Nature
Nanotech.} \textbf{7}, 494--498 (2012).

\bibitem{xiaodongcui} H. Zeng, J. Dai, W. Yao, D. Xiao, \& X. Cui, \textit{%
Nature Nanotech.} \textbf{7}, 490--493 (2012). 

\bibitem{xu1} S. Wu, \textit{et al.} \textit{Nature Phys.} \textbf{9},
149-153 (2013).

\bibitem{xu2} J. S. Ross, \textit{et al.} \textit{Nature Commun.} \textbf{4}%
, 1474 (2013).


\bibitem{Geim} K. S. Novoselov, A. K. Geim, S. V. Morozov, D. Jiang, Y.
Zhang, S. V. Dubonos, I. V. Grigorieva, and A. A. Firsov, Science 306, 666
(2004).

\bibitem{yaowang} D. Xiao, W. Yao and Q. Niu, Phys. Rev. Lett. 99, 236809
(2007).

\bibitem{beenakker} A. Rycerz, J. Tworzydlo and C. W. J. Beenakker, Nature
Physics 3, 172 (2007).

\bibitem{white} D. Gunlycke and C. T. White, Phys. Rev. Lett. 106, 136806
(2011).

\bibitem{Lixiao} X. Li, F. Zhang and Q. Niu, Phys. Rev. Lett. 110,
066803(2013). 


\bibitem{haizhou} H.-Z. Lu, W. Yao, D. Xiao, and S.-Q. Shen, Phys. Rev.
Lett. 110, 016806 (2013).

\bibitem{Asgari} H. Rostami, A. G. Moghaddam and R. Asgari, Phys. Rev. B 88,
085440 (2013).

\bibitem{Andor-prb} A. Kormanyos, V. Zolyomi, N. D. Drummond, P. Rakyta, G.
Burkard, and V. I. Falko, Phys. Rev. B 88, 045416 (2013).

\bibitem{Zhouli} Zhou Li and J. P. Carbotte, Phys. Rev. B 86, 205425 (2012).

\bibitem{Liuguibin} G. Liu, W. Shan, Y. Yao, W. Yao, \& D. Xiao, Phys. Rev.
B 88, 085433 (2013).

\bibitem{Chang} M.-C. Chang and Q. Niu, Phys. Rev. B 53, 7010 (1996).

\bibitem{prx-dot} A. Kormanyos, V. Zolyoml, N. D. Drummond, and G. Burkard,
Phys. Rev. X 4, 011034(2014).

\bibitem{Cai} T. Cai et al, Phys. Rev. B 88, 115140 (2013). 

\bibitem{Kim-prl} Y. Zhang et al, Phys. Rev. Lett. 96, 136806 (2006).

\bibitem{Ong-prl} J. G. Checkelsky, L. Li, and N. P. Ong, Phys. Rev. Lett.
100, 206801 (2008).

\bibitem{Andrei-nat} X. Du, I. Skachko, F. Duerr, A. Luican, and E. Y.
Andrei, Nature 462, 192 (2009).

\bibitem{Kim-natphys} A. F. Young et al, Nature Phys. 8, 550 (2012).

\bibitem{Si-natphy} S. Goswami \textit{et al}, Nature Physics 3, 41 (2007). 

\bibitem{AlAs-prl} Y. P. Shkolnikov, E. P. De Poortere, E. Tutuc, and M.
Shayegan, Phys. Rev. Lett. 89, 226805 (2002).







\bibitem{Rose} F. Rose, M. O. Goeribig and F. Piechou, Phys. Rev. B 88,
125438 (2013). 

\bibitem{Mccann} M. Koshino, E. McCann, Phys. Rev. B 81, 115315 (2010).

\bibitem{Fuchs} J. N. Fuchs and P. Lederer, Phys. Rev. Lett. 98, 016803
(2007).

\bibitem{Aoki-prb} T. Morimoto, Y. Hatsugai, and H. Aoki, Phys. Rev. B 78,
073406 (2008).

\bibitem{Aoki-prl} T. Morimoto, Y. Hatsugai, and H. Aoki, Phys. Rev. Lett.
103, 116803 (2009).
\end{thebibliography}
\end{document}